\RequirePackage{fix-cm}
\documentclass[twocolumn,epjc3]{svjour3}  
\smartqed  
\RequirePackage{graphicx}
\RequirePackage{latexsym}
\RequirePackage[numbers,sort&compress]{natbib}
\RequirePackage[colorlinks,citecolor=blue,urlcolor=blue,linkcolor=blue]{hyperref}
\journalname{Eur. Phys. J. E}
\usepackage{mwe}
\usepackage{widetext}

\begin{document}

\title{Brownian particles driven by spatially periodic noise
}


\author{Davide Breoni\thanksref{e1,addr1}
        \and
        Ralf Blossey\thanksref{addr2} 
        \and
        Hartmut L\"owen\thanksref{addr1}
}

\thankstext{e1}{e-mail: breoni@hhu.de}


\institute{Institut f\"ur Theoretische Physik II: Weiche Materie, Heinrich
Heine-Universit\"at D\"usseldorf, Universit\"atsstra{\ss}e 1, 
40225 D\"usseldorf, Germany \label{addr1}
           \and
           University of Lille, UGSF CNRS UMR8576, 59000 Lille, France \label{addr2}
}

\date{Received: date / Accepted: date}

\abstractdc{We discuss the dynamics of a Brownian particle under the influence of a spatially periodic noise
strength in one dimension using analytical theory and computer simulations. In the absence of a deterministic force, 
the Langevin equation 
can be integrated formally exactly. We determine the short- and long-time behaviour of the mean 
displacement (MD) and mean-squared displacement (MSD). 
In particular we find a very slow dynamics for the mean displacement, scaling as $t^{-1/2}$ with time $t$. Placed under an 
additional external periodic force near the critical tilt value we compute the stationary current obtained from the corresponding Fokker-Planck 
equation and identify an essential singularity if the minimum of the noise strength is zero. Finally, in order to further elucidate the effect of the random periodic driving on the diffusion 
process, we introduce a phase factor in the spatial noise with respect to the external periodic force and identify the value of the phase shift for which the random force exerts its strongest 
effect on the long-time drift velocity and diffusion coefficient.}

\maketitle

\section{Introduction}
Dating back to the important paper by Einstein in the {\it annus mirabilis} 1905 \cite{einstein_uber_1905}, the dynamics of
Brownian particles has been in the focus of statistical physics for more than 100 years now \cite{frey_brownian_2005}.
The constant interest in Brownian particles is basically inspired by two facts: First, their stochastic description requires
fundamental principles such as the Langevin or Smoluchowski picture such that they serve as
paradigmatic models which can be made systematically more complex. Second, there is a variety
of excellent realizations of Brownian particles including mesoscopic colloidal particles in suspension \cite{hansen_liquids_1991},
 random walkers in the macroscopic world (such as \cite{xiong_pedestrian_2012}) and in the microscopic biological context \cite{codling_random_2008},
and even elements of the stock exchange market \cite{tsekov_brownian_2013}. This facilitates
a direct comparison  of the stochastic averages between the stochastic modelling and real experimental data.\\

In its simplest one-dimensional form, the most basic model Langevin equation
for a particle trajectory $x(t)$ as a function of time $t$ is
$\dot{x}(t) = \sqrt{D}\eta(t)$
in which  $\eta(t)$ is white noise with zero mean and variance
$\langle \eta(t) \eta(t')\rangle = \delta(t-t')$ and $D>0$ is the diffusion constant. Here,
$\langle ... \rangle$ denotes a noise average.
With the initial position
$x(t=0)=x_0$, the mean displacement vanishes due to symmetry, $\langle x(t) - x_0 \rangle = 0$, and the mean-squared displacement
is purely diffusive, $\langle (x(t) - x_0)^2 \rangle = 2Dt$. Clearly, this basic equation can be extended towards more complicated situations
including an additional static external force, time-dependent external forcing,
higher spatial dimensions,  and many interacting particles, see  \cite{hess_generalized_1983,nagele_dynamics_1996,dhont_introduction_1996,hanggi_artificial_2009} for some reviews. \\ 
One particularly interesting way to extend the equation is to generalize it to a situation of multiplicative noise, where the noise strength is a positive function $D(x,t)$. While the case where $D$ is only an explicit function of time $t$ is well studied, for example in the context of Brownian ratchets \cite{rousselet_directional_1994,kula_brownian_1998,oudenaarden_brownian_1999,wu_near-field_2016,reimann_brownian_2002} and heat engines \cite{cedraschi_zero-point_2001,kay_synthetic_2007,seifert_stochastic_2012,martinez_brownian_2016,martinez_colloidal_2017}, in this work we focus on the case where we have a {\it spatially dependent noise strength} \cite{buttiker_transport_1987,landauer_motion_1988,van_kampen_explicit_1988,malgaretti_confined_2013} 
 modelled by a positive function $D(x)$, i.e. a space-dependent diffusion coefficient,
such that the most basic model for such processes is given by the Langevin
equation
\begin{equation} \label{langevin}
\dot{x}(t) = \sqrt{D(x(t))}\eta(t) .
\end{equation}
The special case of multiplicative noise where $\dot{x}(t) = -\kappa x(t)\eta(t)$ with
positive $\kappa$ \cite{volpe_effective_2016}, which is somehow related to this model,
documents already that the spatial dependence of the noise gives rise to fundamentally new mathematical
concepts also known as the Itô-Stratonovich problem \cite{mannella_ito_2012}.
The mathematical difficulties associated with the formal treatment of Eq.(\ref{langevin})
are subject to intense discussion, see, e.g., the recent work by Leibovich and Barkai for the specific choice of $D(x)$
as a power-law \cite{leibovich_infinite_2019} and numerous other studies  \cite{lancon_drift_2001,pesek_mathematical_2016,farago_connection_2016,bray_random_2000,pieprzyk_spatially_2016,berezhkovskii_communication_2017,kaulakys_modeling_2009,aghion_infinite_2020,dos_santos_critical_2020,xu_heterogeneous_2020,ray_space-dependent_2020,li_particle_2020,breoni_active_2021,malgaretti_confined_2013}.\\

In this paper we consider a variant of this model in the context of the discussion of particle motion in tilted potentials. There is
a large literature on this topic, see \cite{reimann_giant_2001,reimann_diffusion_2002,sasaki_diffusion_2005,reimann_weak_2008,
evstigneev_diffusion_2008,evstigneev_interaction-controlled_2009,cheng_long_2015,juniper_microscopic_2015,guerin_universal_2017,bai_diffusion_2018,
bialas_colossal_2021,zarlenga_trapping_2007,zarlenga_complex_2009,zarlenga_transient_2019}.
Following the original suggestion by Büttiker \cite{buttiker_transport_1987} and Landauer \cite{landauer_motion_1988} 
the spatially-varying thermal noise source can be combined with a ratchet potential, as e.g. recently discussed by \cite{mazzitello_new_2019}.
Our model considers overdamped Brownian particles subject to an oscillating tilted potential and a space-dependent periodic noise 
amplitude with the same wave vector $k$ as the force; furthermore, we will ultimately also allow a shifted phase $\phi$ in the random force. 
In its general form, the model is given by the Langevin equation
\\
\begin{equation} \label{model}
\gamma \dot{x}(t) = -\nabla V(x)+\sqrt{2 \gamma k_B T(x) }\eta(t),
\end{equation}
where $V(x):=-F_0\left(x + \epsilon \sin(kx)/k \right)$ is the potential, $T(x):=T_0\left(1+\nu \cos(kx+\phi)\right)^2$ is the space-dependent noise strength,  $\gamma$ is the friction coefficient, $F_0$ is the tilting force, $T_0$ is a reference temperature, $\eta(t)$ is a white noise, as introduced before, 
and $\epsilon$ and $\nu$ are dimensionless parameters.  The critical tilt in this model arises when $\epsilon = 1$. In order keep the noise strength differentiable everywhere and its phase in a fixed frame 
 we consider $0\leq\nu\leq 1$. The period of both the force and the noise will be $L=2\pi/k$. We remark that 
the case $\nu=1$ plays a special role insofar as there are special positions at which the noise is zero. In absence of forces, the particle 
will therefore never cross these positions but stay confined within a periodicity length $L$.\\

Our goal in this paper is to describe the particle dynamics as functions of $\epsilon$, $\nu$ and $\phi$, either in the vicinity of the critical tilt, 
or in the absence of the deterministic force, $F_0 = 0$, i.e. in the purely spatial random noise case. Among our main results are the very slow dynamics in the relaxation of the mean displacement (MD) and mean-squared displacement (MSD) for long times in the $F_0=0$ case and an essential singularity in the stationary current for $F_0\neq 0$ and $\epsilon\simeq\nu\simeq 1$. In the case of the full model, we build upon the results of \cite{buttiker_transport_1987} by also considering extreme temperature oscillations where the noise strength vanishes ($\nu=1$) and adding an external driving force, while we expand on \cite{reimann_giant_2001} by finding a theoretical approximation for both the long-time drift $v_L$ and diffusion constant $D_L$ and the phase value $\phi$ for which we have the largest increase of $v_L$ and $D_L$ for $\epsilon\neq 1$ and $\nu\neq 0$. 
Our results have been obtained both with numerical and analytical methods.\\
 
The paper is organized as follows: in the beginning we focus on the free case, for which we study the short- and long-time behaviour of MD and MSD, then we proceed with the full model, including the tilted potential, for which we study the stationary distribution and the dependence of long time diffusion and drift on $\phi$ and $\nu$. Finally, we summarize the results obtained and discuss possible experimental realizations of the model.

\section{Free particle case}

In the case of a vanishing external force ($F_0=0$), the Langevin equation (\ref{model}) now reads as
\begin{equation}
\label{free}
\gamma \dot{x}(t) = \sqrt{2\gamma k_B T(x) } \eta(t),
\end{equation}
where we set $\phi =0$ without loss of generality. We decided to approach this problem using the Stratonovich interpretation. For a given representation of the noise, 
this equation can  be solved by direct integration in the particular case of 
{\it periodic boundary conditions} (PBC) in which we identify $x(t)\pm L$ with $x(t)$. 
The PBC correspond to a ring-like geometry of the one-dimensional system.
\begin{eqnarray}
\label{freesol}
x(t)&=& \frac{2}{k}\arctan\left[\sqrt{\frac{1+\nu}{1-\nu}}\tan \left(k\sqrt{\frac{ k_B T_0(1-\nu^2)}{2\gamma}}\right.\right.\nonumber\\
&&\hspace*{-.5cm}\left.\left.\times\int_0^t\eta(t')dt'+\arctan\left(\sqrt{\frac{1+\nu}{1-\nu}}\tan\left(\frac{kx_0}{2}\right)\right)\right)\right]\nonumber\\
\end{eqnarray}
and the limit of this solution for $\nu\rightarrow 1$ is
\begin{eqnarray}
\label{simplified}
x(t)&=&\frac{2}{k}\arctan\left[k\sqrt{\frac{2 k_B T_0}{\gamma}}\int_0^t\eta(t')dt'+\tan\left(\frac{kx_0}{2}\right)\right] .\nonumber\\
\end{eqnarray}
We remark here that in the case with {\it no boundaries}, i.e. when we let the particle diffuse through the whole $x$ axis, 
the analysis is harder 
and we were not able to find an analytical expression except for the special case $\nu=1$.
In this limit PBC and the no boundaries case are identical  
as the particle can never trespass the points where the noise is zero.\\

Equations (\ref{freesol}) and (\ref{simplified}) can be used to express noise-averages of any power of displacement. For 
an arbitrary moment $M_n(t):= \langle(x(t)-x_0)^n\rangle$ we obtain
\begin{widetext}
\begin{eqnarray}
\label{Mnnu}
M_n(t)& = \int_{-\infty}^{\infty}\left\{\frac{2}{k}\arctan\left[\sqrt{\frac{1+\nu}{1-\nu}}\tan \left(k\sqrt{\frac{ k_B T_0(1-\nu^2)}{2\gamma}}W+\arctan\left(\sqrt{\frac{1+\nu}{1-\nu}}\tan\left(\frac{kx_0}{2}\right)\right)\right)\right] - x_0 \right\}^n \frac{\textrm{e}^{-\frac{W^2}{2t}}}{\sqrt{2\pi t}} dW\nonumber\\ 
\end{eqnarray}
for $\nu\neq 1$ and
\begin{eqnarray}
\label{Mn}
M_n(t)& = \int_{-\infty}^{\infty}\left\{ \frac{2}{k}\arctan\left[k\sqrt{\frac{2 k_B T_0}{\gamma}}W+\tan\left(\frac{kx_0}{2}\right)\right]- x_0 \right\}^n \frac{\textrm{e}^{-\frac{W^2}{2t}}}{\sqrt{2\pi t}} dW  
\end{eqnarray}
for $\nu=1$. Since we are going to focus on the mean displacement $\langle x(t)-x_0\rangle$ and the mean-squared 
displacement $\langle (x(t)-x_0)^2\rangle$, we write the expressions for these two moments $(n=1,2)$ explicitly:
\begin{eqnarray}
\label{MD1nu}
\langle x(t)-x_0\rangle& = \int_{-\infty}^{\infty}\frac{2}{k}\arctan\left[\sqrt{\frac{1+\nu}{1-\nu}}\tan \left(k\sqrt{\frac{ k_B T_0(1-\nu^2)}{2\gamma}}W+\arctan\left(\sqrt{\frac{1+\nu}{1-\nu}} \tan\left(\frac{kx_0}{2}\right)\right)\right)\right] \frac{\textrm{e}^{-\frac{W^2}{2t}}}{\sqrt{2\pi t}} dW - x_0\nonumber\\ 
\end{eqnarray}
and
\begin{eqnarray}
\label{MD2nu}
\langle (x(t)-x_0)^2\rangle& = \int_{-\infty}^{\infty}\left\{\frac{2}{k}\arctan\left[\sqrt{\frac{1+\nu}{1-\nu}} \tan \left(k\sqrt{\frac{ k_B T_0(1-\nu^2)}{2\gamma}}W+\arctan\left(\sqrt{\frac{1+\nu}{1-\nu}}\tan\left(\frac{kx_0}{2}\right)\right)\right)\right] - x_0\right\}^2 \frac{\textrm{e}^{-\frac{W^2}{2t}}}{\sqrt{2\pi t}} dW\nonumber\\ 
\end{eqnarray}
\end{widetext}
for $\nu\neq 1$ and
\begin{eqnarray}
\label{MD1}
\langle x(t)-x_0\rangle & = & \int_{-\infty}^{\infty}\frac{2}{k}\arctan\left[k\sqrt{\frac{2 k_B T_0}{\gamma}}W\right.\nonumber\\
&&\left.+\tan\left(\frac{kx_0}{2}\right)\right] \frac{\textrm{e}^{-\frac{W^2}{2t}}}{\sqrt{2\pi t}} dW - x_0 \nonumber\\
\end{eqnarray}
and
\begin{eqnarray}
\label{MD2}
\langle (x(t)-x_0)^2\rangle & = & \int_{-\infty}^{\infty}\left\{\frac{2}{k}\arctan\left[k\sqrt{\frac{2 k_B T_0}{\gamma}}W\right.\right.\nonumber\\
&&\left.\left.+\tan\left(\frac{kx_0}{2}\right)\right]-x_0\right\}^2 \frac{\textrm{e}^{-\frac{W^2}{2t}}}{\sqrt{2\pi t}} dW \nonumber\\
\end{eqnarray}
for $\nu=1$.

\subsection{Short-time behavior}

We can use equations (\ref{MD1nu}-\ref{MD2}) to extract the short-time behavior of  the MD and MSD. 
Expanding the integrand in powers of $t$ using a Taylor series and integrating the 
terms separately we obtain for the MD:
\begin{eqnarray} \label{12}
\langle x(t)-x_0\rangle &=&-\frac{ k k_B T_0 t}{\gamma}\sin(k x_0)\nu\nonumber\\
&&\hspace{-1.5cm}\times\left[(1+\nu)\cos^2\left(\frac{k}{2}x_0\right)+(1-\nu)\sin^2\left(\frac{k}{2}x_0\right)\right]\nonumber\\
&&\hspace{-1.5cm}+\mathcal{O}\left(t^2\right)
\end{eqnarray}
and for the MSD
\begin{eqnarray} \label{13}
\langle (x(t)-x_0)^2\rangle=\frac{2 k_B T_0 t}{\gamma}\left[(1+\nu)\cos^2\left(\frac{k}{2}x_0\right)\right.\nonumber\\
\left. + (1-\nu)\sin^2\left(\frac{k}{2}x_0\right)\right]^2+\mathcal{O}\left(t^2\right).
\end{eqnarray}
In the special limit $\nu=1$ we also add the second order correction as:
\begin{eqnarray} \label{14}
 \langle x(t)-x_0\rangle & = & -\frac{2 k k_B T_0 t}{\gamma}\cos^2\left(\frac{k}{2}x_0\right)\sin(kx_0)\nonumber \\
&& \times \left[1
 -\frac{6k^2k_B T_0 t}{\gamma}\cos^2\left(\frac{k}{2}x_0\right)\cos(k x_0)\right]\nonumber\\
&& +\mathcal{O}\left(t^3\right)
\end{eqnarray}
and
\begin{eqnarray} \label{15}
 \langle (x(t)-x_0)^2\rangle &=& \frac{8 k_B T_0 t}{\gamma}\cos^4\left(\frac{k}{2}x_0\right)\nonumber \\
&&\hspace*{-1.5cm}\times\left[1+\frac{k^2k_B T_0 t}{\gamma}\cos^2\left(\frac{k}{2}x_0\right)\left(7-11\cos\left(kx_0\right)\right)\right]\nonumber\\
&&\hspace*{-1.5cm} +\mathcal{O}\left(t^3\right).
\end{eqnarray}
Clearly, the first-order correction of  (\ref{14}) and (\ref{15}) coincides with 
 equations (\ref{12}) and (\ref{13}) in the limit $\nu\to1$. Moreover for $\nu=0$ we recover the 
white noise case solved by Einstein \cite{einstein_uber_1905}.

\begin{figure}[!htbp]
 \begin{center}
 \includegraphics[width=0.45\textwidth]{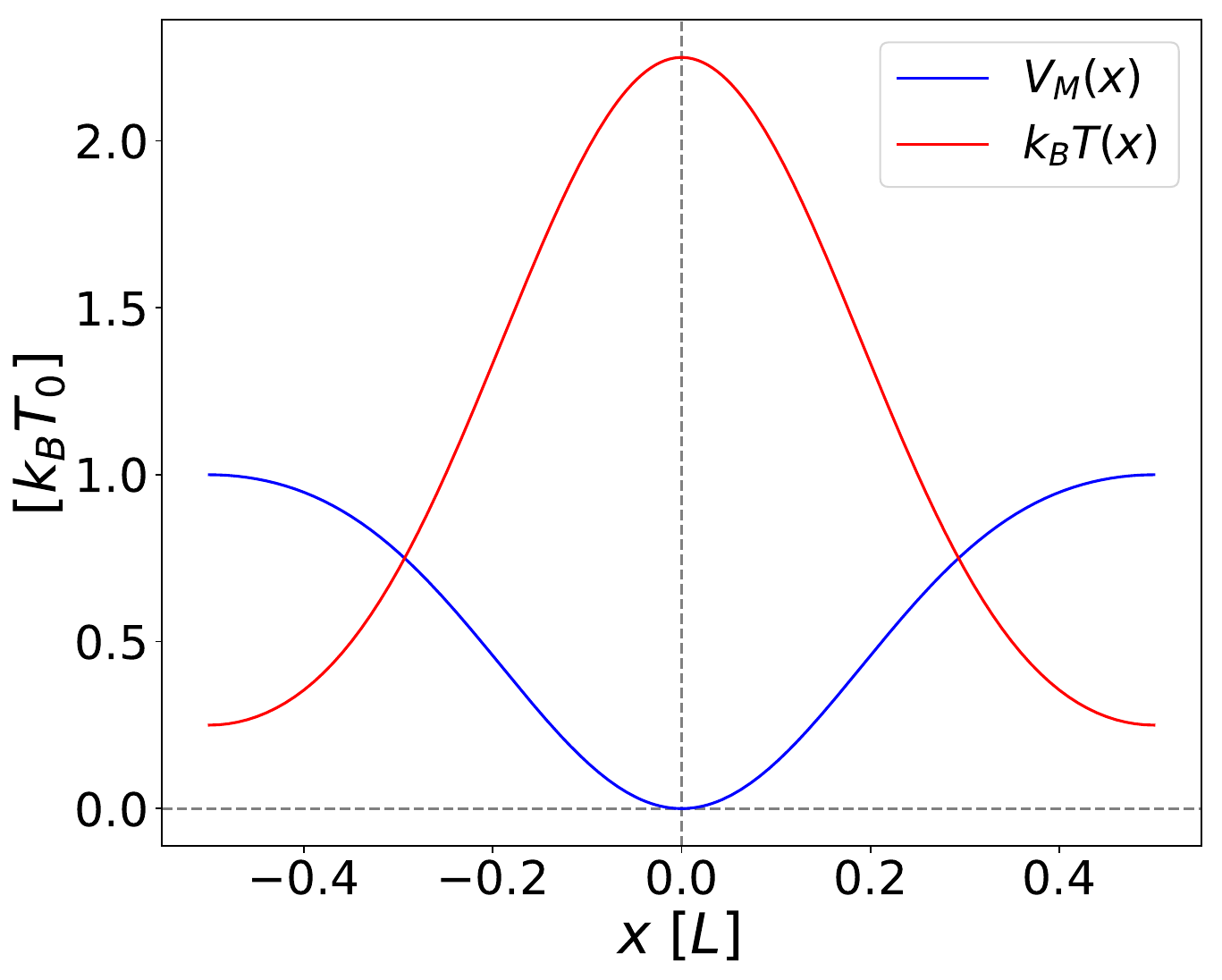}
 \caption{Effective potential of the mean displacement $V_M$, obtained from the short-time drift of the mean displacement, and space-dependent noise $T(x)$ for $\nu = 0.5$ and $\phi=0$ as functions of space $x$. While the averaged MD tends to the minima of $V_M$,
 where the noise strength $T(x)$ is largest, individual trajectories spend most of their time around the maxima of $V_M$.}
 \label{Figure1}
\end{center}
\end{figure}

We now define an {\it effective potential of the mean displacement} such that a particle subject to this potential and constant white noise will experience the same average drift as a particle in a space-dependent noise landscape. In other words, following the spirit of the mapping proposed by Büttiker \cite{buttiker_transport_1987}, the effective force resulting from this potential can be viewed 
as a substitute source for the drift when only white noise is considered. 
Hence we define this force $F_M(x)$  up to a friction coefficient prefactor $\gamma$
as the first coefficient of the short-time expansion of the MD
\begin{equation}
\label{16}
\langle x(t)-x_0\rangle=a_1(x_0,\nu)t+\mathcal{O}\left(t^2\right) 
\end{equation}
as follows
\begin{eqnarray}
\label{17}
 F_M(x) & := & a_1(x,\nu)\gamma\nonumber\\
 & = & -k k_B T_0 \sin(k x)\nu\left[(1+\nu)\cos^2\left(\frac{k}{2}x\right)\right.\nonumber\\
 &&\left.+(1-\nu)\sin^2\left(\frac{k}{2}x\right)\right] .
\end{eqnarray}
The effective potential of the mean displacement is then defined by $V_M(x)=-\int_0^xF_M(x')dx'$ yielding
\begin{eqnarray}
\label{18}
V_M(x)&=&k_B T_0 \nu \nonumber\\
&&\hspace{-1.5cm}\times\left[(1-\nu)\sin^4\left(\frac{k}{2}x\right)-(1+\nu)\cos^4\left(\frac{k}{2}x\right)+1+\nu\right].\nonumber\\
\end{eqnarray}
This potential is shown in Fig. \ref{Figure1}. Even though this potential is defined just by the short-time expansion of the MD, it is still significant for any finite time, as the particle is overdamped and feels at every time a short-time drift depending only on its position. As a result, the MD of a particle subject to this potential and white noise can be perfectly mapped to the MD of a free particle with space-dependent noise.\\

While the average mean displacement behaves according to $V_M$, moving over time towards the regions 
where $V_M$ is smaller and the noise strength is larger, we want to stress that individual trajectories 
will not accumulate in the minima of $V_M$ but will instead freely move over all the domain, spending most of their 
time in the maxima of $V_M$ instead. This is because when particles reach such low-noise regions they take a longer 
time escaping, as their fluctuations there are severely reduced.

\subsection{Dynamics for finite and long times}

Now we explore the behavior of the MD and MSD for finite and long times. First we 
present an asymptotic analysis for the special case $\nu=1$. Then we use 
a numerical solution of the integrals in (\ref{MD1nu}) and (\ref{MD2nu}) as well as 
computer simulations of the original Langevin equation 
to obtain data for finite times and arbitrary $\nu$.

\subsubsection{Asymptotic analysis for $\nu=1$ for long times}
Here we present an asymptotic analysis for the MD and MSD by starting from Eq.(\ref{MD1}) and using
 the asymptotic approximation 
\begin{equation}
\label{19}
\arctan(\theta)\simeq \frac{\pi}{2}\textrm{sign}(\theta)-\arctan\left(\frac{1}{\theta}\right),
\end{equation}
for large $\theta$. We now expand $\arctan\left(\frac{1}{\theta}\right)$ using Euler's formula \cite{chien-lih_8967_2005}
\begin{equation} \label{20}
\arctan\left(\frac{1}{\theta}\right)=\frac{\theta}{\theta^2+1}+\mathcal{O}\left(\frac{1}{\theta^3}\right)
\end{equation}
and insert this expansion in  Eq.(\ref{MD1}) to obtain
\begin{widetext}
\begin{eqnarray} \label{21}
\langle x(t)-x_0\rangle & = & \int_{-\infty}^{\infty}\frac{2}{k}\left[ \frac{\pi}{2}\textrm{sign}\left(k\sqrt{\frac{2 k_B T_0t}{\gamma}}W+\tan\left(\frac{kx_0}{2}\right)\right)-\frac{k\sqrt{\frac{2 k_B T_0t}{\gamma}}W+\tan\left(\frac{kx_0}{2}\right)}{1+\left(k\sqrt{\frac{2 k_B T_0t}{\gamma}}W+\tan\left(\frac{kx_0}{2}\right)\right)^2}\right]\frac{\textrm{e}^{-\frac{W^2}{2t}}}{\sqrt{2\pi t}} dW\nonumber\\
&&- x_0+\mathcal{O}\left(t^{-3/2}\right), 
\end{eqnarray}
which yields
\begin{eqnarray} \label{22}
\langle x(t)-x_0\rangle &=& \sqrt{\frac{\pi \gamma}{k_BT_0 t}}\frac{1}{k^2}\tan\left(\frac{kx_0}{2}\right)\left(1-\frac{1}{2k}\sqrt{\frac{\gamma}{k_BT_0t}}\right) -x_0+\mathcal{O}\left(t^{-3/2}\right).
\end{eqnarray}
\end{widetext}
As a result, the leading asymptotic  behavior of $\langle x(t)\rangle$ is determined by the first term involving a scaling 
behavior
of the MD in $1/\sqrt{t}$. This is remarkably slow compared to typical behavior 
of a Brownian particle in a harmonic potential or of active 
Brownian motion where the MD reaches its asymptotic value exponentially in time 
\cite{howse_self-motile_2007,ten_hagen_brownian_2011,sprenger_time-dependent_2021} 
thus constituting an example of a very slow relaxation as induced by space-dependent noise.

Likewise an asymptotic analysis for $\nu=1$ yields for the long-time limit of the MSD 
\begin{equation} \label{23}
\lim_{t\rightarrow \infty}\left(\langle (x(t)-x_0)^2\rangle\right) =x_0^2+\frac{\pi^2}{k^2}
\end{equation}
which represents the degree of smearing of the particle distribution for long times. We want to remark that the MSD calculated from a distribution with periodic boundary conditions does not describe the effective diffusion coefficient $D_L$ in periodic systems with no boundaries, in contrast to the MD which can actually be calculated from the distribution with periodic boundary conditions even for open systems.
\\

\subsubsection{Computer simulations}
We performed direct Brownian dynamics computer simulations of the original Langevin equations with a finite time step $\Delta t$
to obtain numerically results for the MD and MSD at any times. In order to properly simulate a system with space-dependent noise,
 we used the order $\mathcal{O}\left(\Delta t\right)$ Milstein scheme \cite{milshtejn_approximate_1975} 
 with a time step of $\Delta t=10^{-3}\tau$, where $\tau:= \frac{\gamma L^2}{k_BT_0}$ is a 
 typical Brownian time scale of the system. For each simulation set we fixed the initial position $x_0$ within 
 the first period $[-L/2,L/2]$ and averaged typically over 200 trajectories of length $\simeq 500\tau$.

\subsubsection{MD and MSD for finite times}

\begin{figure*}[!htbp]
 \begin{center}
 \includegraphics[width=\textwidth]{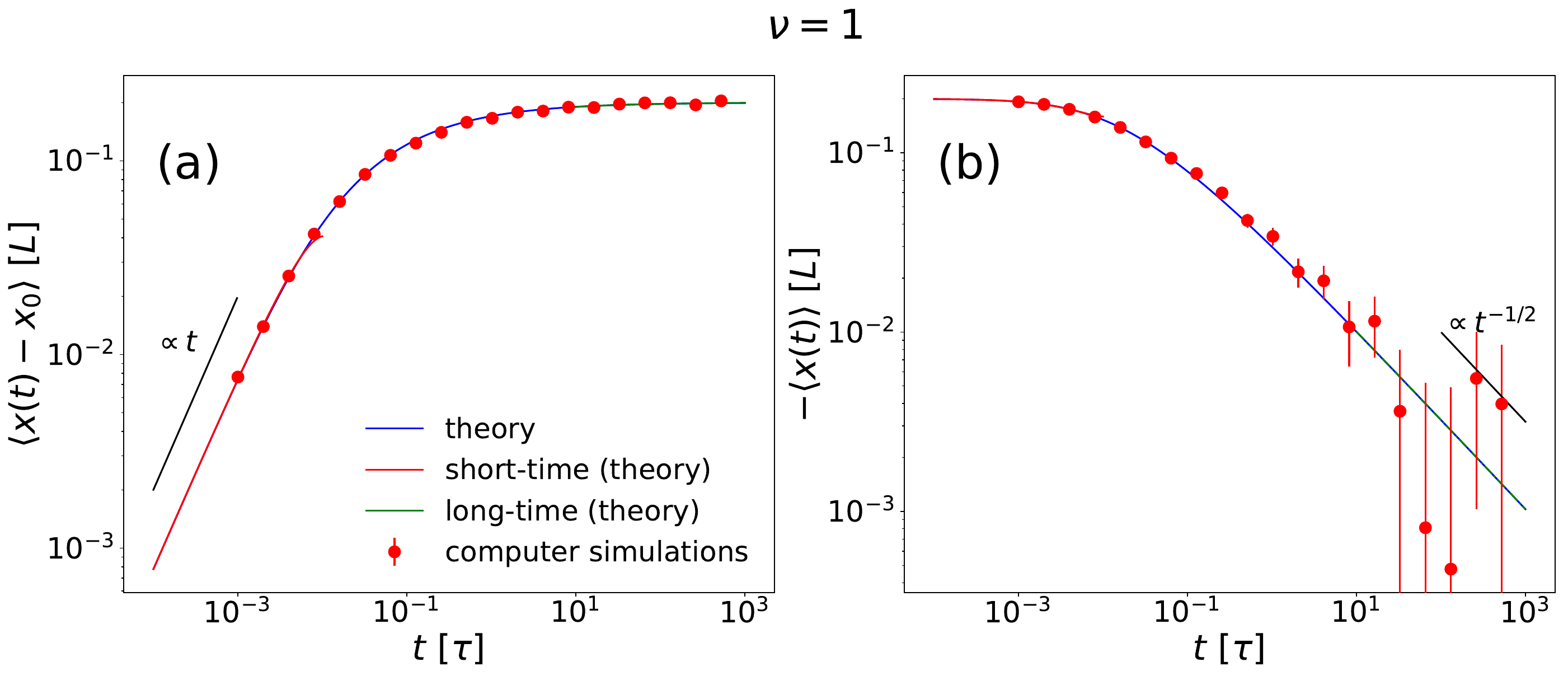}
 \caption{Absolute values of the mean displacement $\langle x(t)-x_0 \rangle$ (a) and the mean position 
 with a minus sign to ensure positivity $-\langle x(t)\rangle$ (b) 
 for $\nu=1$ and $x_0=-0.2L$ as a function of time $t$. The numerical evaluation of the integral 
 in Eq.\ (\ref{MD1}) (theory) and its asymptotic short- and long-time expansions (\ref{14}) and (\ref{21})
 are shown together with simulation data.  The MD increases linear in time  $t$ for short times, while the decay to its limit 
 scales in a very slow way with $\mathcal{O}\left(t^{-1/2}\right)$.}
 \label{Figure2}
\end{center}
\end{figure*}
\begin{figure*}[!htbp]
 \begin{center}
 \includegraphics[width=\textwidth]{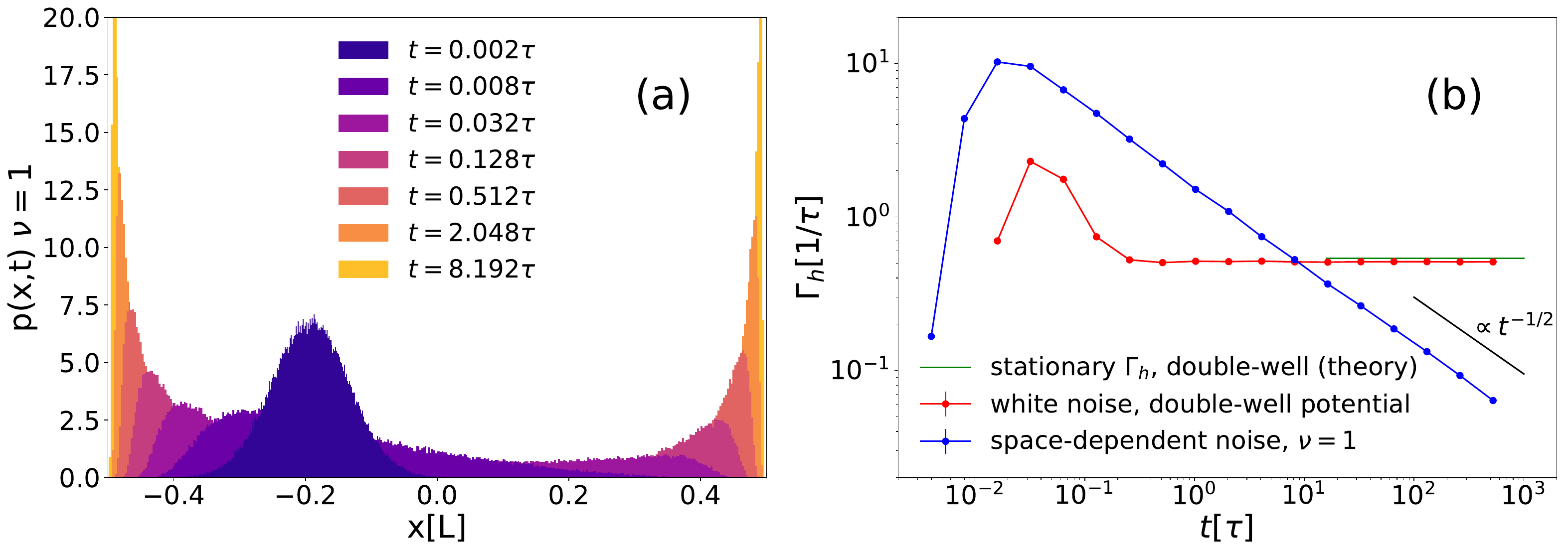}
 \caption{(a) Probability density function $p(x,t)$ for the particle position at different times $t$, 
 with $\nu=1$ and $x_0=-0.2L$. Here we averaged over 10000 different trajectories of length $10\tau$. (b) Hopping rate $\Gamma_h$ between the two peaks in the particle distribution as a function of time $t$
for a space-dependent noise with points of vanishing noise ($\nu=1$), and for a double-well potential with white noise. Here we have chosen $x_0=-0.2L$. We also show the stationary state theoretical value of $\Gamma_h$ for the double well potential, defined as the inverse of the mean first passage time $t_e$, derived in Eq.(\ref{rate}). }
\label{Figure3}
\end{center}
\end{figure*}

Data for the mean displacement and the mean position 
 as a function of time are obtained by a numerical evaluation of the integral 
 in Eq.\ (\ref{MD1}) and by computer simulation. For $\nu=1$ results are presented in Fig. \ref{Figure2} together with the 
 corresponding short-time and long-time asymptotics (\ref{14}) and (\ref{22}). The displacement starts linear in time $t$ and
 saturates for long times. The mean position approaches zero slowly as a power law in time proportional to $t^{-1/2}$.
 For large times the statistical error in the simulation data is significant but nevertheless these data are 
 compatible with the scaling prediction of the theory.

In order to understand the very slow behavior of the MD we note that 
while the MD tends to zero, i.e. to the point with largest noise, 
this is just an effect of averaging over particles spending most of their time at the points 
with the smallest noise on both sides of the $x$-axis: $ x \simeq -L/2 $ and $ x \simeq L/2$.
 This particular mechanism explains why the MD approaches its final value so slowly, as the particles have to hop from one side to the other to symmetrize their distribution. In  Fig. \ref{Figure3}a this is clearly documented in the time evolution of the particle distribution function $p(x,t)$, which gives the probability to find a particle after a time $t$ at position $x$ provided it started at time $t=0$ at position $x_0$.  The system evolves from a single-peaked distribution around $x_0$ 
 to a double-peaked distribution in $\pm L/2$. Near the two points $x=\pm L/2$ of zero noise the peaks are getting sharper
as $t\to\infty$ approaching to $\delta$-peaks such that $\lim_{t \to \infty}p(x,t)= (\delta( x-L/2) + \delta( x+L/2))/2$.
The intuitive reason for this is that once a particle adsorbs at the points $x=\pm L/2$ of zero noise it will never return to the region where the noise is finite.\\
This peculiar behavior is clearly delineated from the relaxation in a symmetric double-well potential with white noise of strength $T_0$.
In order to reveal this, we have performed simulations for a Brownian particle 
in the double-well potential with two equal minima 
\begin{equation}
U(x):=A(x^4 - Bx^2).
\end{equation} 
We set $A:=48k_BT_0/L^4$ and $B:=L^2/2$ in order to have the two wells in $\pm L/2$ such that the energy barrier between the two minima is $3k_BT_0$. Our simulation for this white-noise reference case show  that both the MD and the MSD decay exponentially in time $t$ 
rather than with $1/\sqrt{t}$, and hence much faster than for our case of space-dependent noise. 
We also defined a particle  hopping rate $\Gamma_h$ between the two peaks of the distribution as
\begin{equation}\label{24}
\Gamma_h(t):=\frac{N_h(t)}{t/2},
\end{equation}
where $N_h(t)$ is the number of times a particle hops from one peak to the other in the time interval $[t/2,t]$. Note that the relevant time window in which hopping is considered is chosen to be proportional in time in order to improve the statistics. We have a {\it hop} whenever the particle trespasses the $x=L/4$ or $x=-L/4$ thresholds and previously was respectively in the left or right peak.\\
 In fact, as we show in Fig. \ref{Figure3}b, for the double-well potential, the hopping rate $\Gamma_h(t)$ 
 converges to a constant for long times. This rate is maintaining the equilibrium state with a symmetrized occupation around the two minima. The rate saturates for $t\to \infty$ to a value very close to the inverse of the mean first passage time (see for example \cite{malgaretti_active_2022}) in the double-well potential $t_e$ \cite{caprini_correlated_2021}, which in our case is given by:
\begin{eqnarray}\label{rate}
 t_e&\simeq & \frac{2\pi}{\sqrt{\nabla^2U(L/2)|\nabla^2U(0)|}}\exp{\left(\frac{U(0)-U(L/2)}{k_BT_0}\right)}\nonumber\\
 &\simeq & 1.859\tau.
\end{eqnarray}  
Conversely, for our case of space-dependent noise, the hopping rate keeps decreasing as a function of time again with an inverse power law $t^{-1/2}$. This reflects the fact that
 the peaks of the space-dependent noise distribution keep growing indefinitely as the particles get in average closer to the points of zero noise.

\begin{figure*}[!htbp]
 \begin{center}
 \includegraphics[width=\textwidth]{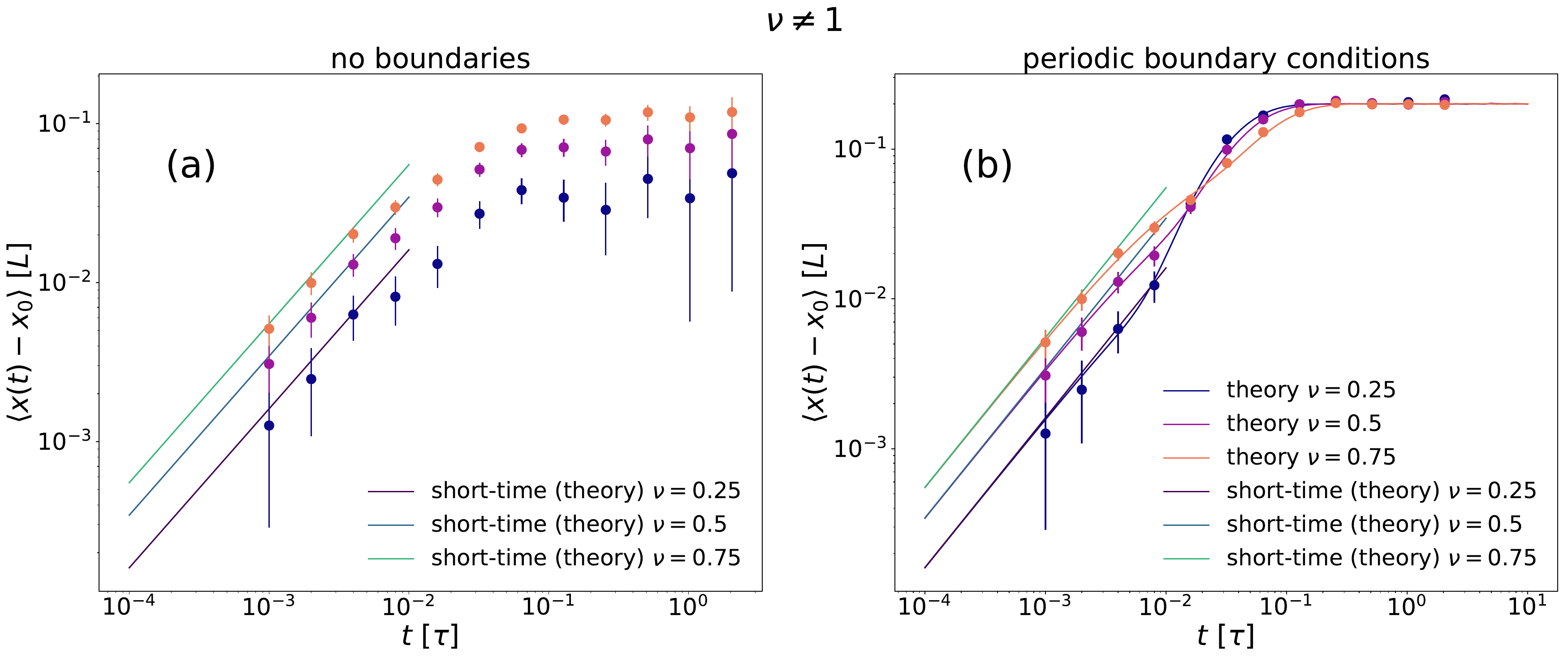}
\caption{Mean displacement $\langle x(t)-x_0 \rangle$ (a,b) as a function of time $t$
 for three values of  $\nu=0.25,0.5,0.75$ and $x_0=-0.2L$  
 for  no boundaries (a) and for periodic boundary conditions (b).}
\label{Figure4}
\end{center}
\end{figure*}

Now in Fig. \ref{Figure4} we explore the MD for the case $\nu \not= 1$
 where the particle crosses the position of minimal noise. 
Here the boundary conditions do matter and we distinguish between no boundaries
 (Fig. \ref{Figure4}a) with infinitely many oscillations and periodic
 boundary conditions of a ring-like geometry (Fig. \ref{Figure4}b). 
While the short-time behavior is linear in time for both kind of boundary conditions, 
the MD saturates for long times to a finite value depending on $\nu$ and $x_0$ for the no boundaries case.
This finite value is $-x_0$ for periodic boundary conditions since in this case the mean position will always end 
at zero due to symmetry. The asymptotic approach to zero is exponential in time
 as in the case of the double-well potential with noise as the particle stays mobile even when approaching the position
where the noise is minimal. This is in marked contrast to the limit of $\nu=1$ where the particle gets immobilized at the boundaries.\\

\begin{figure*}[!htbp]
\begin{center}
\includegraphics[width=\textwidth]{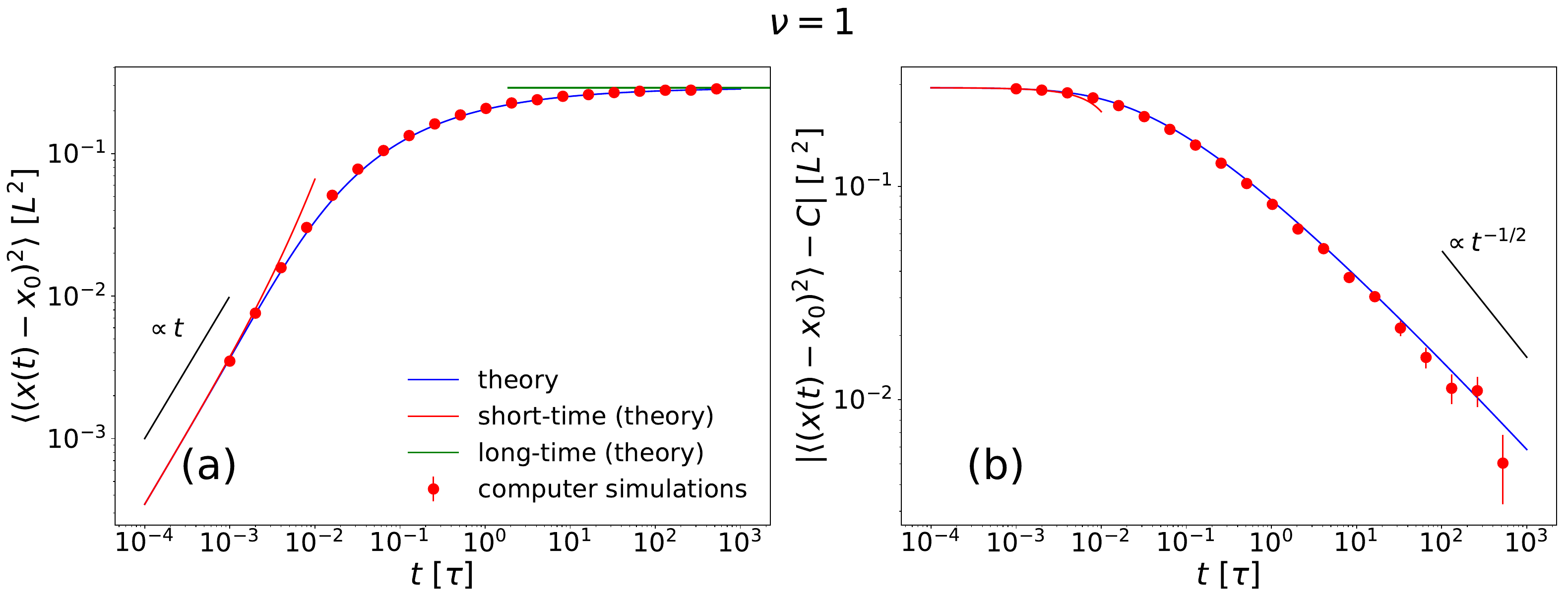}
\caption{
Absolute values of the mean-squared displacement (MSD) $\langle (x(t)-x_0)^2 \rangle$  (a) and shifted
MSD$-C$, where $C$ is the long time limit of the MSD (b) as a function of time $t$
 for $\nu=1$ and $x_0=-0.2L$. The numerical evaluation of the integral in Eq. (\ref{MD2}) and 
 its asymptotic short- and long-time expansions are shown together with simulation data. }
\label{Figure5}
\end{center}
\end{figure*}

Now we turn to the MSD, first for the special case $\nu=1$ shown in Fig.\ \ref{Figure5}a where boundary conditions do not matter. The MSD starts linear 
in time and then saturates to its long-time limit $C:=x_0^2+(\pi/k)^2$. Its asymptotic approach 
to this saturation value is revealed by plotting the MSD shifted by $C$  which decays to zero for large times, see Fig.\ \ref{Figure5}b. Similar to the MD for $\nu=1$,
 we find that the asymptotic behavior is compatible with a $1/\sqrt{t}$ scaling.\\

\begin{figure*}[!htbp]
 \begin{center}
 \includegraphics[width=\textwidth]{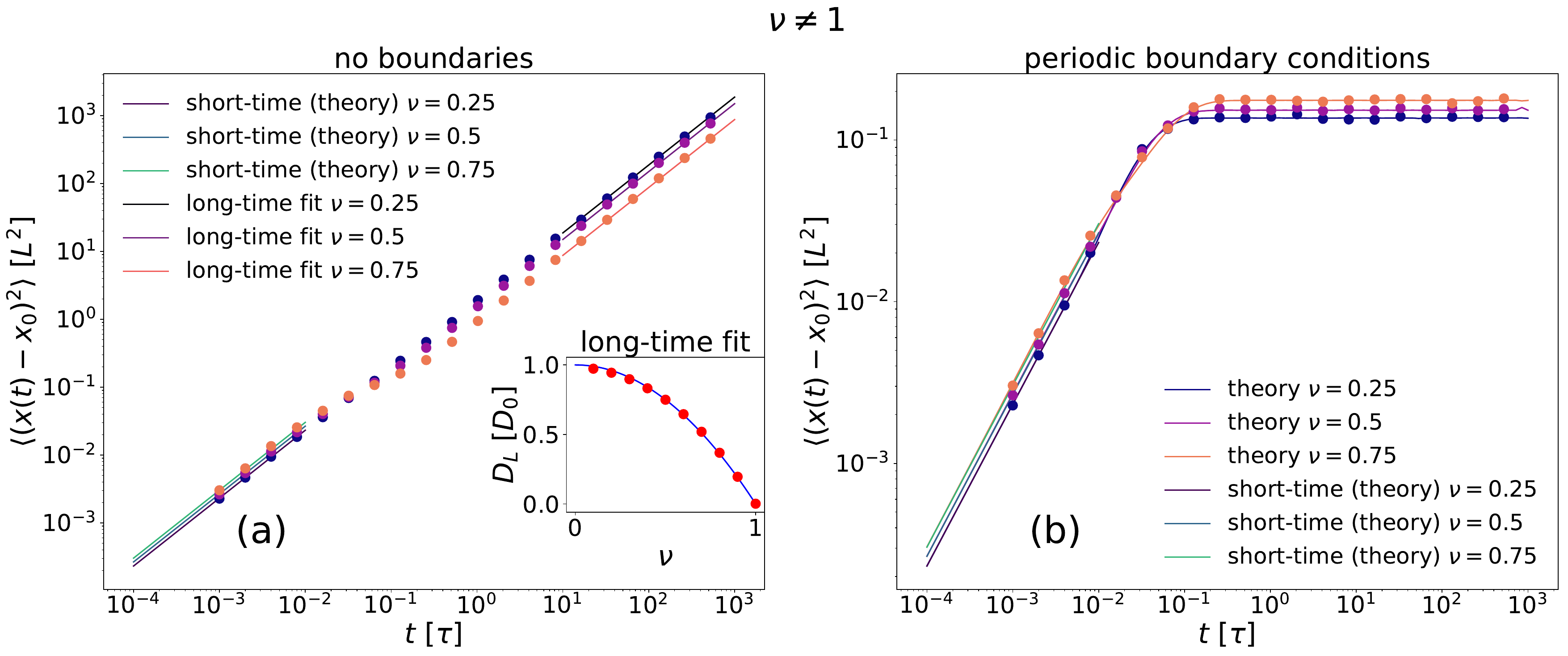}
 \caption{(a) Mean-squared displacement (MSD) $\langle (x(t)-x_0)^2 \rangle$ as a function of time $t$
 for $\nu\neq 1$ and $x_0=-0.2L$ 
 both for no boundaries (a) and for periodic boundary conditions (b). The inset shows 
the long-time diffusion constant $D_L$ as a function of $\nu$ for the no boundaries case.}
 \label{Figure6}
\end{center}
\end{figure*}

In  Fig.\ \ref{Figure6} we show the MSD for $\nu\not=1$ for both types of boundary conditions. 
In absence of boundary conditions (see Fig.\ \ref{Figure6}a)  the long-time behavior is linear in time 
 $\approx 2D_Lt$ involving a long-time diffusion coefficient $D_L$. Clearly
the latter depends on $\nu$ but not on the initial position $x_0$. This dependence is depicted in the inset of 
Fig.\ \ref{Figure6}a. We found the empirical expression $D_L(\nu ) = D_0(1-\nu^2)$ with
$D_0=k_BT_0/\gamma$ to be a very good  fit to the data.
This can be regarded  as a parabolic fit which fulfills the inflection symmetry in $\nu$ and the
constraint $D_L(\nu =1 )=0$. The same behavior was recently found in a similar system \cite{caprini_dynamics_2022}. \\
\begin{figure*}[!htbp]
 \begin{center}
 \includegraphics[width=\textwidth]{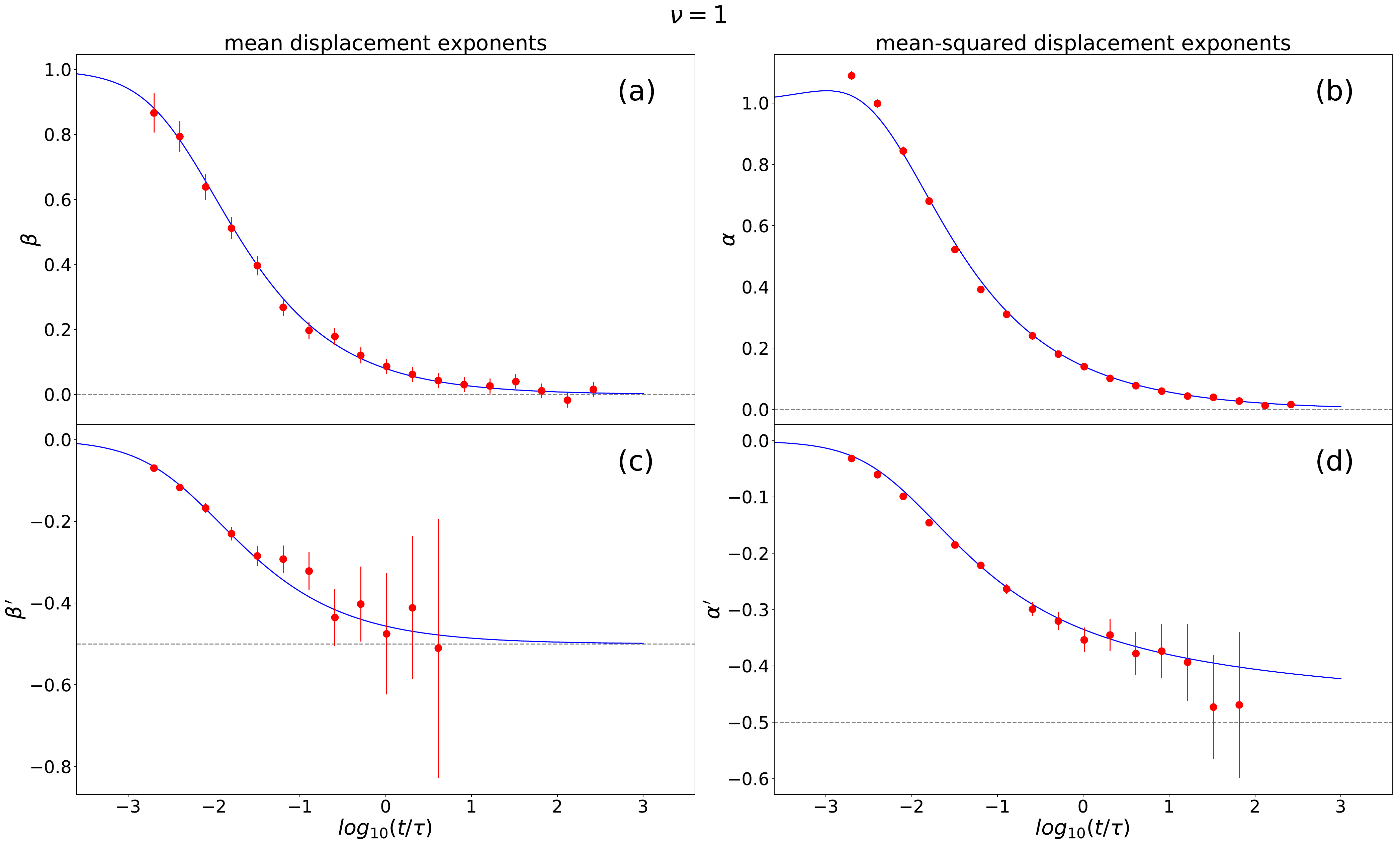}
 \caption{Dynamical exponents of the MD (a,c) and MSD (b,d) close to 0 (a,b) and their final 
 limit (c,d) for $\nu=1$ and $x_0=-0.2L$ in theory and simulation as functions of time $t$. 
 As we have already seen in Fig. \ref{Figure2} 
 for the MD and Fig. \ref{Figure4} for the MSD, both quantities grow initially linearly in time 
 and decay to their final limit with $1/\sqrt{t}$ for the MD and slower than $1/\sqrt{t}$ for the MSD.}
 \label{Figure7}
\end{center}
\end{figure*} 
Finally, to better clarify the behaviors of the MD and MSD for $\nu=1$, we plot the dynamical exponents (Fig.\ref{Figure7}) that define the scaling regimes for the MD 
($\beta$, $\beta '$) and MSD ($\alpha$, $\alpha '$) close to their short-time and long-time limits, respectively:
\begin{eqnarray} \label{25}
\beta &:= & \frac{d( \log_{10}|\langle x(t)-x_0\rangle|)}{d(\log_{10}(t))}\, ,\,\,\, \nonumber\\
\beta' &:= & \frac{d( \log_{10}|\langle x(t)\rangle|)}{d(\log_{10}(t))},\nonumber\\
\alpha &:= & \frac{d( \log_{10}\langle (x(t)-x_0)^2\rangle)}{d(\log_{10}(t))}\,,\,\,\,\nonumber\\
  \alpha'&:= & \frac{d( \log_{10}|\langle (x(t)-x_0)^2\rangle-C|)}{d(\log_{10}(t))}.\nonumber\\
\end{eqnarray}
Both the MD and MSD for short times are linear, while for long times the scaling of the MD converges clearly to -0.5, 
that corresponds to $1/\sqrt{t}$. Within the time window explored the MSD has not yet saturated to an ultimate 
dynamical exponent
 for long times. The asymptotics shown is compatible with a final scaling exponent of $-1/2$ 
 although the approach to this final exponent is much slower for the MSD than for the MD where the 
 saturation is clearly visible. 

We remark that an  algebraic asymptotic approach in the MSD was also found 
for equilibrium Brownian dynamics of repulsive interacting particles. 
Here the time-derivative of the time-dependent diffusion coefficient
MSD$/t$ scales as  $t^{-d/2}$ in $d$ spatial dimensions
\cite{cichocki_time-dependent_1991,ackerson_correlations_1982,lowen_structure_1992,kollmann_single-file_2003} but the physical origins of the algebraic scaling laws are different.

\section{Tilted potential} 

\begin{figure}[!htbp]
\begin{center}
\includegraphics[width=.45\textwidth]{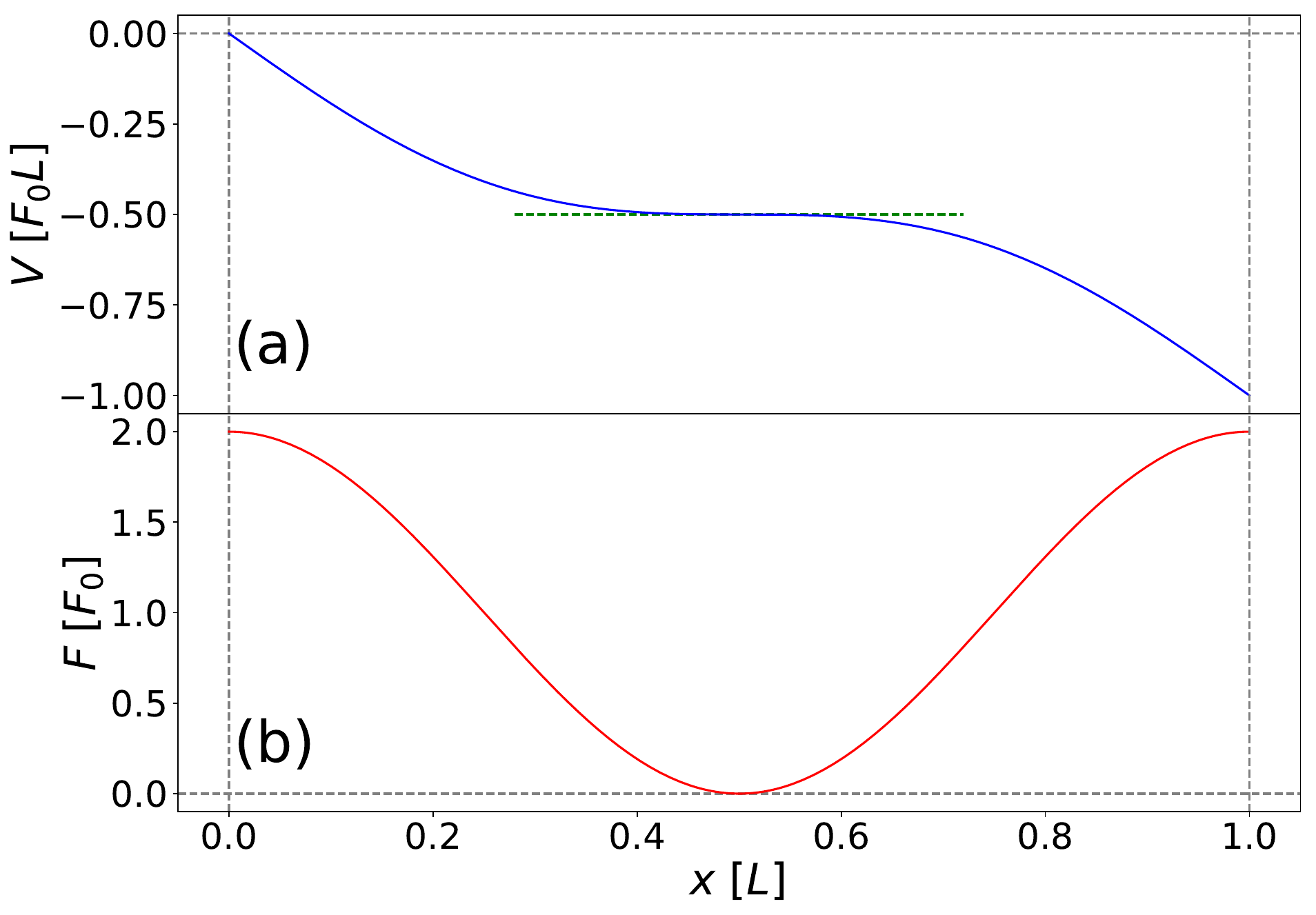}
\caption{Potential (a) and corresponding force (b) near the `flat' regime $\epsilon = 1$ 
as a function of the spatial coordinate $x$.}
\label{Figure8}
\end{center}
\end{figure}

In this section we leave the situation in which the Brownian particle is a free particle only driven by spatially-dependent noise. We now consider
the full model, including the deterministic tilted potential. We first look at the situation near the critical value of the amplitude $\epsilon = 1$, where the
tilted potential develops a plateau. The situation addressed in shown in Fig. \ref{Figure8}.

\subsection{The stationary current}

Being weakly confined to a region of the deterministic potential in which the dynamics can be considered `slow', a quasi-stationary
distribution can be defined \cite{guerin_universal_2017}.
The Fokker-Planck equation corresponding to the Langevin equation, Eq. (\ref{langevin}) in Stratonovich interpretation reads as 
\begin{eqnarray}
\partial_t p(x,t)  &=&  - \partial_x a(x) p(x,t) + \frac{1}{2}\partial_x [b(x)\partial_x [b(x)p(x,t)]] \nonumber\\
\end{eqnarray}
with $a(x)$ the force and $b(x)$ the noise amplitude,
\begin{equation}
a(x):= F_0(1+ \epsilon \cos(kx))\,,\,\,\, b(x) := \sqrt{2\gamma k_BT(x)}\, .
\end{equation}
Following the discussion in \cite{guerin_universal_2017}, the dynamics near the critical tilt value for $\epsilon \geq 1$ is 
characterized by a stationary current given by the one-time integrated FP-equation
\begin{equation}
-J_s = - a(x)p_s(x) + \frac{1}{2}b(x)\partial_x[b(x)p_s(x)]\, . 
\end{equation}
Defining $(b(x)/2)p_s(x) = \widehat{p}_s(x) $ we can rewrite the last expression as
\begin{equation}
-\frac{J_s}{b(x)} = - R(x)\widehat{p}_s(x) + \partial_x \widehat{p}_s(x) 
\end{equation}
with
\begin{equation}
R(x) = \frac{2a(x)}{b^2(x)}\,. 
\end{equation}
The equation can be solved with the Ansatz $\widehat{p}_s(x) = u(x)\cdot v(x)$ which reduces the problem to two
readily integrable first-order ordinary differential equations for $u(x)$ and $v(x)$. 
One obtains the final expression
\begin{equation}
p_s(x) = \frac{2J_s}{b(x)} \int_x^{\infty} dy \frac{1}{b(y)} \exp{\left(-\int_x^y dz R(z)\right)}\, ,
\end{equation} 
in which the current $J_s$ can be obtained from the normalization integral $\int_{-\infty}^{\infty} dx p_s(x) = 1$.
In the following we take for simplicity (setting all other constants to one)
\begin{equation}
R(z) = \frac{1 + \epsilon \cos(z)}{(1 + \nu \cos(z))^2}\, .
\end{equation}
Setting $b(z) := \exp\left(-F(z)\right) =  \exp(-\ln(1+ \nu \cos(z)) $, and expanding both $b(z)$ and
$R(z)$ in Taylor series around the center of the flat region near $z = L/2$, the stationary current $J_s$
is given by\\
\begin{eqnarray} 
J_s &=& \frac{1}{2} 
\left[
\int_{-\infty}^{\infty} dx \exp{\left(-\widehat{F}(x)\right)} 
\right.\nonumber \\
 &&\left.\times \int_0^{\infty} dy  
\exp{\left(-\widehat{F}(y)\right)} 
\exp{
\left(-
\int_x^y dz \widehat{R}(z)
\right)
}
\right]^{-1}\nonumber\\ 
\end{eqnarray}
in which the symbol $\widehat{...} $ indicates the Taylor-expanded functions, 
\\
\begin{equation} \label{fz}
\widehat{F}(x) = \ln(1-\nu) + \frac{1}{2}\left(\frac{1}{1-\nu}\right)(x - L/2)^2
\end{equation}
and
\begin{equation} \label{rz} 
\widehat{R}(z) = 2\frac{1 - \epsilon}{(1-\nu)^2} + \frac{\epsilon(\nu +1) - 2\nu}{(1-\nu)^3}(z -L/2)^2\, .
\end{equation}
The integration of $\widehat{R}(z)$ yields a cubic polynomial, but due to cancellations the resulting expression in the
exponential is Gaussian in $x$ and cubic in $y$. The Gaussian integral in $y$ can be calculated exactly,
while the remaining expression in $y$ needs to be evaluated numerically for each value of $\epsilon$ and $\nu$.

The most interesting behavior of the stationary current is found in the limit $\nu \rightarrow 1 $, $\epsilon \approx 1$.
The fact that the coefficients in Eqs.(\ref{fz}),({\ref{rz}) are singular in $1/(1 - \nu)$ leads to a singular behavior of $J_s$
in the form
\begin{equation}
J_s \propto (1-\nu)^{m} \exp\left[-\frac{I(\epsilon,\nu)}{(1-\nu)^{n}}\right]\, ,
\end{equation}
with $n = 3$, since the dominant singularity in $\widehat{R}(z)$ is $\propto (1- \nu)^{-3}$, see Eq.(\ref{rz}). 
The amplitude is $I(\epsilon,\nu) = (\epsilon(\nu +1) - 2\nu)/4$  and the rational factors 
combine to $ m = 0$. The stationary current thus goes to zero with an essential singularity in $ (1 - \nu)$.
 
\subsection{Phase difference between noise and potential} 

For a tilted potential, we now explore the effect of a non-zero phase $\phi \neq 0$ on the long-time behavior 
of the particle for different values of $\nu$ by using computer simulation.\\
As shown in \cite{reimann_giant_2001}, the long-time drift velocity and diffusion coefficients ($v_L$ and $D_L$ respectively) can be 
analytically calculated for the case $\nu=0$, where we set $V(x)$ as potential.\\
Here the question is how the mismatch of the periodic 
noise and external forcing affects the long-time behavior of the particle. Intuitively one would expect that 
overcoming an energetic barrier is best if the maximum of the noise occurs where the external force is opposing most.
Then the noise would help to bring the particle over the energetic barrier. The position where the force is opposing most is clearly given for $x=L/2+nL$, where $n$ is an integer. Then it is expected that mobility gets a maximum if the phase shift is $\phi=\pi$. This is indeed what we confirm by simulation. 
We chose  $k_BT_0=0.01 F_0L$ and  $\epsilon=1.3$. The potential barrier $\Delta E$ is given by
\begin{eqnarray} \label{DE}
\Delta E(\epsilon)&=&\frac{F_0L}{\pi}\left(\sqrt{\epsilon^2-1}-\textrm{arcsec} (\epsilon)\right),
\end{eqnarray}
yielding $\Delta E\simeq 0.04 F_0L > 0.01 F_0L$ for $\epsilon=1.3$. 

Given these parameters, we simulated the system for different values of 
$\phi$ and $\nu$ and results are summarized in Fig. \ref{Figure9}. 
Since to the best of our knowledge there is no easy generalization of the results in \cite{reimann_giant_2001} for a space-dependent temperature, we have compared the simulation data with a mapping on the analytical results for $v_L^{(0)}$ and $D_L^{(0)}$
\cite{reimann_giant_2001} which were obtained for a spatially constant temperature. 
Since the crucial position to hop over the barrier is at $x=L/2+nL$ where the opposing force is maximal, 
this represents the kinetic bottleneck for the dynamical process. Therefore it is tempting to compare
our simulation results with the analytical ones where this local noise strength $T(x=L/2)$ is inserted as a homogeneous temperature. We remark that this temperature $T(x=L/2)$ depends both on the oscillation strength $\nu$ and the phase shift $\phi$ of $T$ with respect to the potential. This mapping theory should work best if the particle spends most of its time close to the point $x=L/2$. In fact,
Fig. \ref{Figure9} reveals that this simple mapping theory describes the simulation data well even for large $\nu$. 
As a function of the phase mismatch $\phi$, both $D_L$ and $v_L$ are enhanced when $\phi$ is between about $\frac{3}{5}\pi$ and $\frac{7}{5}\pi$. Clearly around the value $\phi=\pi$ we find the maximal enhancement of both $v_L$ and $D_L$.
In the complementary case, the noise strength $T(x)$ has its minimum closer to the crucial region where the opposing force is maximal, and as a result the drift velocity and diffusion are severely reduced. 
For $\nu=1$ they are even brought exactly to zero when $|\phi|<\textrm{arcsec}(\epsilon)$, since the particle is stuck and there is no systematic external force to drift over the positions of vanishing noise.
\begin{figure*}[!htbp]
 \begin{center}
 \includegraphics[width=\textwidth]{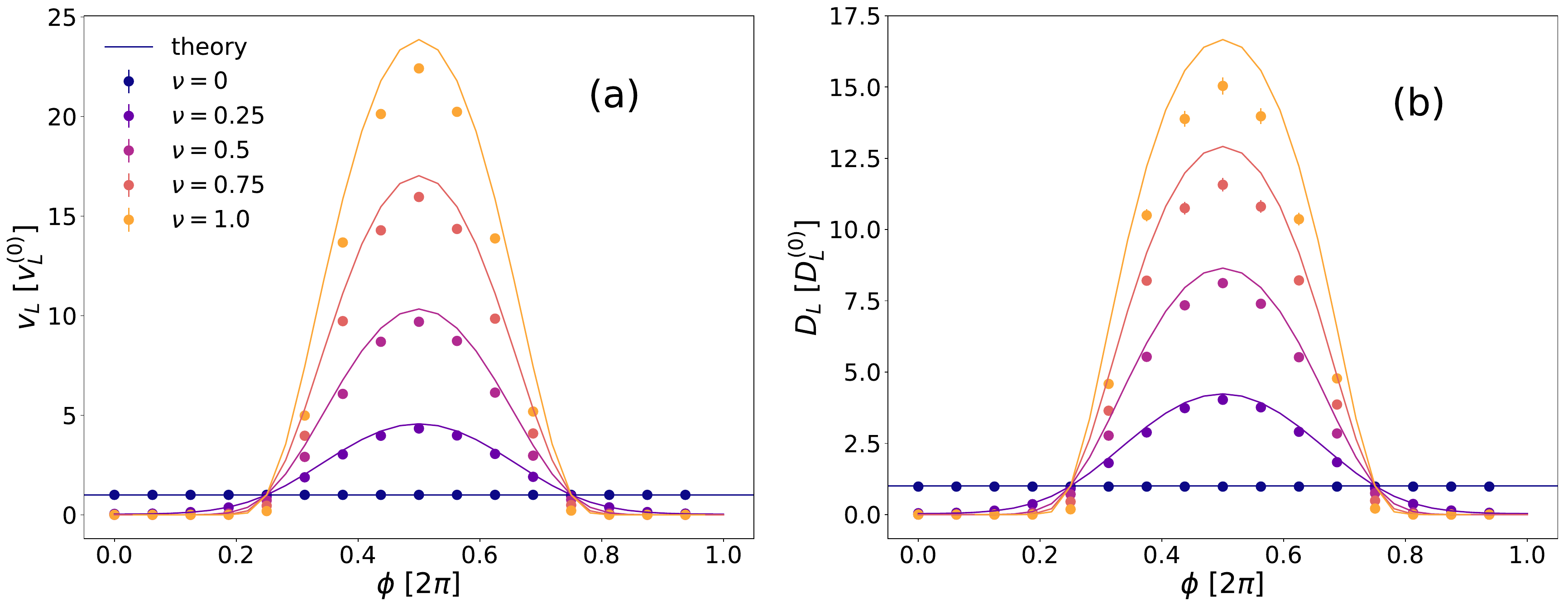}
 \caption{Long time drift (a) and diffusion (b) for $\epsilon=1.3$ as functions of $\nu$ and $\phi$, 
 simulations results and theory. For $\phi$ between about $\frac{3}{5}\pi$ and $\frac{7}{5}\pi$ both $D_L$ and $v_L$ are enhanced, having a maximum in $\phi=\pi$, while otherwise they are reduced.}
 \label{Figure9}
\end{center}
\end{figure*}

\section{Conclusions and outlook}
\label{Sec:conclusions}

In conclusion we have presented a detailed study of a model for a Brownian particle moving in a one-dimen\-sional environment with a space-periodic noise and under an external potential with 
a tilt near its critical value.
In the free case we calculated the exact solution of the associated Langevin equation, and further explicitly obtained short- and long-time approximations of the MD and MSD.
These results allow us to characterize the slow decay of these quantities at long times. Interesting relaxation dynamics occurs around points of vanishing noise which establish centers of growing peaks in the particle distribution, as particles slow down significantly in the neighborhoods of these points. 
Introducing the tilted periodic potential we first determined the stationary current for the quasi-stationary state, which for $\epsilon \geq 1$ displays an essential singularity for the maximal strength of 
the noise oscillations, $\nu$. Finally, we determined numerically the effects of a space-periodic noise on the long-time diffusion and drift as functions of the phase difference between noise and potential $\phi$ and the 
strength of the noise oscillations $\nu$, finding the largest enhancements to take place for a phase of $\phi=\pi$ and the maximal possible noise oscillations for $\nu=1$.
\\

Our one-dimensional model with both periodic boundary conditions or no boundaries can be realized by a colloidal particle
confined in a ring or a linear channel respectively by e.g. optical forces \cite{lutz_single-file_2004,lutz_diffusion_2004,juniper_microscopic_2015,juniper_dynamic_2017}. The space-dependent noise can be added by various means. First, one can change locally the solvent temperature. This realization has a limited applicability, since the state of the solvent can be changed drastically upon such a temperature variation. However, there are more general and more important realizations for our model. First of all, the viscosity or the friction coefficient can directly be changed without changing the ambient temperature. The solvent viscosity, for instance, can be tuned over orders of magnitude by imposed patterned substrates interacting with the solvent or even by varying the size of the colloids without changing the solvent phase. Second, space-dependent noise can stem from active internal fluctuations \cite{frangipane_dynamic_2018,arlt_painting_2018} different from thermal fluctuations and can be embodied into an effective noise strength that can largely be tuned by activity \cite{szamel_self-propelled_2014,wittmann_effective_2017,caprini_active_2018,dabelow_irreversibility_2019,caprini_entropy_2019}. Optical gradients can be used to steer activity as a function of the position, as realized and discussed in \cite{caprini_dynamics_2022,lozano_phototaxis_2016,lozano_propagating_2019,soker_how_2021}. Another possibility is to tune the noise amplitude of {\it skyrmions}, which have a similar equation of motion \cite{brown_effect_2018}. Last but not least, the noise can be mimicked in valuable model systems by applying randomized kicks of an external field to the particle. For example, the noise strength can largely be tuned externally without changing the solvent at all by tuning the rotational diffusion constant of the colloids \cite{fernandez-rodriguez_feedback-controlled_2020,sprenger_active_2020}. In fact, the effective diffusion constant of an active particle depends on its rotational diffusion constant, and in the limit of short persistence lengths one can indirectly tune the translational diffusion by tuning the rotational one.  

\section{Acknowledgements}
We thank L Caprini for confirming the long-time behavior of the MSD described in Fig. \ref{Figure6}a in an analogous system and T Voigtmann for the interesting discussions. 
DB is supported by the EU MSCA-ITN ActiveMatter, (proposal No. 812780). RB is grateful to HL for the invitation to a stay at the Heinrich-Heine University in 
D\"usseldorf where this work was performed. HL was supported by the DFG project LO 418/25-1 of the SPP 2265. 

\section*{Author contribution statement}
HL and RB directed the project. DB performed analytic calculations
and
numerical simulations.
RB contributed analytical results. All authors discussed the results
and
wrote the manuscript.

\providecommand{\newblock}{}

\end{document}